\documentclass[
    ,final            
    ,numberedheadings 
  ]
  {aipproc}

\layoutstyle{6x9}

\begin{document}

\title{Panel Discussion on Scalar Mesons}

\classification{13.25.Ft,14.40.Lb,14.40.Cs,11.80.Et}
\keywords {Scalar mesons; quark model; decay mechanisms; dynamical resoances}

\author{Jonathan L. Rosner}{
  address={Enrico Fermi Institute, University of Chicago,
Chicago, IL 60637}
}
\begin{abstract}
A panel discussion on scalar mesons was held with the participation of David
Bugg, Yulia Kalashnikova, Keh-Fei Liu, Michael Scadron, the author, and
members of the audience.  Some introductory remarks are noted here.
\end{abstract}

\maketitle

\section{Introduction}

I would like to begin by posing several questions for discussion.

\begin{enumerate}

\item What are the masses and widths of whatever scalar $q \bar q$ mesons
fulfill the roles of $^3P_0$ partners of the well-established $^3P_2$ states?

\item What are the $f_0(980)$ and $a_0(980)$?  Are they mainly $q \bar q$,
diquark-antidiquark, meson-meson states, or some mixture?

\item Does a broad $\sigma$ resonance exist in the $I_{\pi \pi}=0$ channel
below 1 GeV?

\item Does a broad $\kappa$ resonance exist in the $I_{K \pi} = 1/2$ channel
below 1 GeV?

\item Are $\sigma$, $\kappa$ related to $f_0(980)$ and $a_0(980)$ by any
symmetry?  Can we thus relate decays of heavier mesons to $SP$ ($S$ = scalar,
$P$ = pseudoscalar) or $S \to PP$ decays to one another?

\item What experiments or analyses would shed further light on these questions?

\end{enumerate}

I would also like to mention some relevant early work on $^3P_0$ meson decays
and dynamical resonance generation in the low-enery $\pi \pi$ system.

\section{Quark-antiquark scalar mesons}

Decays of $^3P_0$ $q \bar q$ mesons may be treated by a single-quark-transition
language \cite{Melosh}, equivalently expressed in several other approaches
\cite{SQT}, relating all D wave decays of $^3P_{1,2}$ mesons to $VP$ and $PP$
to one another, and all S wave decays of $^3P_{1,0}$ mesons to $VP$ and $PP$ to
one another.  $P$ and $V$ stand for pseudoscalar and vector mesons.  Thus
\cite{Colglazier:1970vx}, D waves and S waves in $a_1 \to \rho \pi$ and $b_1
\to \omega \pi$ are related to one another, leading to the prediction
$2(g_1/g_0)_{a_1} = (g_0/g_1)_{b_1} + 1$.

The situation will be complicated by mixing of scalars with $q q \bar q
\bar q$, $PP$ channels.  Nonetheless it is important to understand what
the predictions are for unmixed $q \bar q$ scalars.  The best handle on
the S-wave decays of scalar mesons $S$ to $PP$, using the above language,
is the S-wave amplitude for $a_1 \to \rho \pi$.

\section{Dynamical $\pi \pi$ resonance generation}

The dynamics of pion-pion scattering near threshold is captured by the current
algebra description of Weinberg \cite{Weinberg:1966kf}.  One then wishes to
unitarize such amplitudes.  This is equivalent to summing bubble diagrams to
account for rescattering.  As unitarization in the $s$ channel destroys
crossing symmetry, one must restore it in some manner.  One way to do this
(approximately) is using dispersion relations to construct amplitudes with
appropriate singularities.

An approach in Ref.\ \cite{Brown:1971ea} obtains interesting results in the
limit of massless pions.  If a $\rho$ pole is present in the $I=J=1$ channel,
its mass is ${\cal O}(2 \pi f_\pi)$ and its $\pi \pi$ coupling is predicted
Other consequences are automatic generation of a broad $\sigma$ near the
$\rho$ mass, and the ability to describe rescattering in non-elastic processes
such as $\gamma \gamma \to \pi^+ \pi^-$ \cite{Goble:1972rz}, $\gamma \gamma \to
\pi^0 \pi^0$ \cite{Goble:1988cg}, and $K_L \to \pi^0 \pi^+ \pi^- \to \pi^0
\gamma \gamma$ \cite{Ko:1989is}.

The dynamical generation of low-enery $\pi \pi$ resonances may have an echo
at a mass scale about 2650 = $v/f_\pi$ times as high, where $v = 246$ GeV is
the Higgs boson vacuum expectation value, in the possible dynamical generation
of a Higgs boson in $WW, ZZ$ scattering \cite{Lee:1977yc,Hung:1984gw}.  We
look forward to experiments at the CERN Large Hadron Collider (LHC) to tell
us whether our experience from years ago with low-energy pion dynamics may
stand us in good stead at the TeV scale.

\begin{theacknowledgments}
 I thank the organizers of this Workshop for providing a
stimulating environment for discussion.  This work was supported in part
by the United States Department of Energy under Grant No.\ DE FG02 90ER40560. 
\end{theacknowledgments}


\begin{thebibliography}{99}

\bibitem{Melosh}
  H.~J.~Melosh,
  \emph{Phys.\ Rev.\ D} {\bf 9} (1974) 1095;
  F.~J.~Gilman, M.~Kugler and S.~Meshkov,
  \emph{Phys.\ Lett.\ B} {\bf 45}, 481 (1973);
  \emph{Phys.\ Rev.\ D} {\bf 9}, 715 (1974);
  F.~J.~Gilman and I.~Karliner,
  \emph{Phys.\ Rev.\ D} {\bf 10}, 2194 (1974);
  A.~J.~G.~Hey and J.~Weyers,
  \emph{Phys.\ Lett.\ B} {\bf 44}, 263 (1973).

\bibitem{SQT}
  L.~Micu,
  \emph{Nucl.\ Phys.\ B} {\bf 12}, 239 (1969);
  F.~Buccella, H.~Kleinert, C.~A.~Savoy, E.~Celeghini and E.~Sorace,
  \emph{Nuovo Cim.\ } {\bf 69A}, 133 (1970);
  E.~W.~Colglazier and J.~L.~Rosner,
  \emph{Nucl.\ Phys.\ B} {\bf 27}, 349 (1971);
  R.~P.~Feynman, M.~Kislinger and F.~Ravndal,
  \emph{Phys.\ Rev.\ D} {\bf 3}, 2706 (1971);
  W.~P.~Petersen and J.~L.~Rosner,
  \emph{Phys.\ Rev.\ D} {\bf 6}, 820 (1972);
  A.~Le Yaouanc, L.~Oliver, O.~Pene and J.~C.~Raynal,
  \emph{Phys.\ Rev.\ D} {\bf 8}, 2223 (1973);
  \emph{ibid.} {\bf 9}, 1415 (1974).

\bibitem{Colglazier:1970vx}
  E.~W.~Colglazier and J.~L.~Rosner,
  \emph{Nucl.\ Phys.\ B} {\bf 27}, 349 (1971).

\bibitem{Weinberg:1966kf}
  S.~Weinberg,
  \emph{Phys.\ Rev.\ Lett.\ } {\bf 17}, 616 (1966).

\bibitem{Brown:1971ea}
  L.~S.~Brown and R.~L.~Goble,
  \emph{ Phys.\ Rev.\ D} {\bf 4}, 723 (1971).

\bibitem{Goble:1972rz}
  R.~L.~Goble and J.~L.~Rosner,
  \emph{Phys.\ Rev.\ D} {\bf 5}, 2345 (1972).

\bibitem{Goble:1988cg}
  R.~L.~Goble, R.~Rosenfeld and J.~L.~Rosner,
  \emph{Phys.\ Rev.\ D} {\bf 39}, 3264 (1989).

\bibitem{Ko:1989is}
  P.~Ko and J.~L.~Rosner,
  \emph{Phys.\ Rev.\ D} {\bf 40}, 3775 (1989).

\bibitem{Lee:1977yc}
  B.~W.~Lee, C.~Quigg and H.~B.~Thacker,
  {\emph Phys.\ Rev.\ Lett.\ } {\bf 38}, 883 (1977);
  Phys.\ Rev.\  D {\bf 16}, 1519 (1977).

\bibitem{Hung:1984gw}
  P.~Q.~Hung and H.~B.~Thacker,
  {\emph Phys.\ Rev.\ D} {\bf 31}, 2866 (1985).

\end{thebibliography}
\end{document}